\documentstyle[aps,eqsecnum,floats]{revtex}
\begin{document}
\draft
\twocolumn[\hsize\textwidth\columnwidth\hsize\csname 
           @twocolumnfalse\endcsname
\title{Radiative falloff in Einstein-Straus spacetime}
\author{William G.~Laarakkers and Eric Poisson}
\address{Department of Physics, University of Guelph, Guelph,
         Ontario, Canada N1G 2W1}
\date{Submitted to Physical Review D, May 3, 2001} 
\maketitle
\begin{abstract}
The Einstein-Straus spacetime describes a nonrotating black hole
immersed in a matter-dominated cosmology. It is constructed by
scooping out a spherical ball of the dust and replacing it with a
vacuum region containing a black hole of the same mass. The metric is  
smooth at the boundary, which is comoving with the rest of the
universe. We study the evolution of a massless scalar field in the
Einstein-Straus spacetime, with a special emphasis on its late-time
behavior. This is done by numerically integrating the scalar wave
equation in a double-null coordinate system that covers both portions
(vacuum and dust) of the spacetime. We show that the field's evolution
is governed mostly by the strong concentration of curvature near the
black hole, and the discontinuity in the dust's mass density at the
boundary; these give rise to a rather complex behavior at late
times. Contrary to what it would do in an asymptotically-flat
spacetime, the field does not decay in time according to an inverse
power-law.  
\end{abstract}
\vskip 2pc]

\narrowtext

\section{Introduction} 

The dynamics of radiative fields in spacetimes containing a
black hole has been the subject of many investigations, some
analytical \cite{1,2,3,4,5,6,7,8,9}, some 
numerical \cite{10,11,12,13,14,15}, some restricted to
nonrotating black holes \cite{1,2,3,4,5,6,10,11}, and some devoted to
rotating black holes \cite{7,8,9,12,13,14,15}. Most of these studies
were concerned with black holes immersed in an asymptotically-flat
spacetime. In such cases, radiative dynamics always proceeds in the
same three stages. First, an outburst of radiation is emitted, as
determined by the initial conditions imposed on the radiative field;
most of this radiation proceeds directly to infinity, with only mild
distortions produced by the spacetime curvature. Second, the
radiation begins to oscillate, with frequencies and damping times
determined by the properties of the black hole; these oscillations are
produced by the part of the initial outburst that does not go directly
to infinity, but interacts with the high concentration of curvature
near the black hole. Third, the oscillations stop, and the field
decays monotonically, as an inverse power-law in time; the power is 
determined by the radiation's multipole order $l$.  

A number of authors have studied radiative dynamics in
black-hole spacetimes that are not asymptotically flat. Brady,
Chambers, Krivan, and Laguna \cite{16} considered the evolution of a
massless scalar field in Reissner-Nordst\"om-de Sitter spacetime,
which describes a charged black hole immersed in an exponentially
expanding universe. They found that while the first two stages of
radiative dynamics are not affected by the different conditions at
infinity, the third one is: at late times the field decays
exponentially, not as an inverse power-law. This new type of
behavior was explored more thoroughly by Brady, Chambers, Laarakkers,
and Poisson \cite{17}, who showed that the decay constant can be
complex if the scalar field is nonminimally coupled to the 
curvature: the field oscillates as it decays exponentially. While
these authors were concerned with black-hole spacetimes with a
positive cosmological constant, other authors have examined radiative
dynamics in the presence of a negative cosmological
constant \cite{18,19,20,21}; they also find an exponential decay.  

The literature reviewed in the preceding paragraph indicates that
inverse power-law decay is a property of asymptotically-flat
spacetimes, and that different behaviors should be expected in 
black-hole spacetimes that have different asymptotic properties. This 
conclusion has been reinforced by a number of analytical 
studies \cite{3,6}. Our goal with this paper is to contribute further
to the exploration of radiative dynamics in black-hole spacetimes that
are not asymptotically flat. Here we consider the evolution of a
massless scalar field in the spacetime of a nonrotating black hole
immersed in a matter-dominated cosmology. We investigate the effect of
the cosmological asymptotics on the late-time behavior of the
radiation.  

To construct the spacetime we start with a homogeneous, isotropic, 
dust-filled universe and we remove from it a spherical ball of the
dust, which we replace by a vacuum region containing a black hole of
the same mass. The boundary between the vacuum and the dust is
comoving with the rest of the universe. Its dynamics is determined by
the curvature of the spatial sections: it expands forever for zero and
negative curvature, and it eventually recontracts for positive
curvature. This construction was first carried out by Einstein and
Straus in 1945 \cite{22}. We reproduce it in Sec.~II, where we also
introduce a single coordinate system to cover both regions (vacuum and
dust) of the spacetime. In Sec.~III we write down the second-order
partial differential equation that governs the evolution of a massless
scalar field $\Phi$ in this spacetime. After separation of the
variables, the field equation reduces to a two-dimensional wave
equation with an effective potential. This wave equation is
numerically integrated in Sec.~IV, where we present our main results.    

Radiative dynamics in the Einstein-Straus spacetime is governed mostly 
by the strong concentration of curvature near the black hole, and the
discontinuity in the dust's mass density at the boundary. These
features are encoded in the effective potential, which possesses a
fairly well localized barrier near the black hole's event horizon, and
a jump discontinuity at the boundary. A wave propagating in the
presence of this potential will be partially transmitted and reflected
each time it encounters the barrier or the jump. These processes give
rise to a rather complex late-time behavior for the radiation --- 
the field does not decay according to an inverse power-law. We
describe this behavior in detail in Sec.~IV. In Sec.~V we seek insight
into the multiple transmissions and reflections that take place
between the event horizon and the boundary by working through a simple
toy model. 

We believe that radiative dynamics in the Einstein-Straus spacetime
provides an excellent illustration of the general fact that inverse
power-law falloff can occur only in asymptotically-flat spacetimes. 
This article contributes to an ongoing effort to explore the rich
spectrum of behaviors that results when asymptotic flatness is
replaced by different asymptotic conditions. 

Throughout the paper we work in relativistic units, in which 
$c = G = 1$.        

\section{Einstein-Straus spacetime} 

The Einstein-Straus spacetime \cite{22} represents a nonrotating black
hole immersed in a cosmological universe containing a pressureless
fluid (which we shall call dust). It is constructed by scooping out a 
spherical ball of the dust and replacing it with a vacuum region
containing a black hole of the same mass. The mathematical description
of the spacetime involves two metrics joined at a common boundary
$\Sigma$. The metric inside the boundary is given by the Schwarzschild
solution, the metric outside is one of the
Friedmann-Lema\^{\i}tre-Robertson-Walker (FLRW) metrics, and $\Sigma$
is a three-dimensional hypersurface generated by timelike geodesics of
both spacetimes. 

The boundary is comoving with the universe; it expands forever if the
universe has spatial sections that are flat ($k=0$) or have negative
curvature ($k=-1$), or it expands to a maximum radius and recontracts
if the spatial sections have positive curvature ($k=1$). As seen from
the Schwarzschild side, the boundary is either gravitationally bound
to the black hole (if $k=1$), or it is unbound or just marginally
bound (if $k=-1$ or $k=0$, respectively). The spacetime is depicted,
by means of a conformal diagram, in Fig.~1; in this figure the
boundary is assumed to expand forever.   

\begin{figure}
\special{hscale=33 vscale=33 hoffset=-10.0 voffset=0.0
         angle=-90.0 psfile=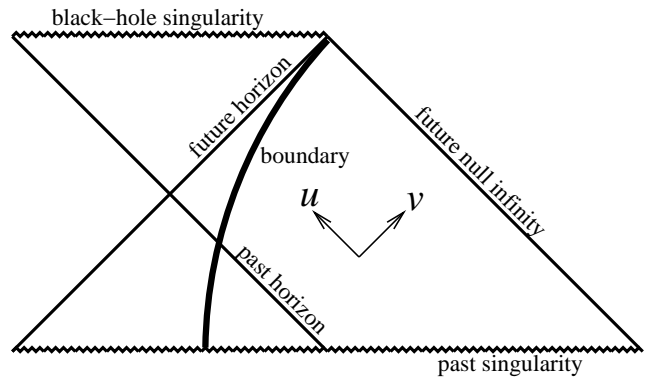}
\vspace*{2.5in}
\caption{The Einstein-Straus spacetime, as represented by a conformal 
  diagram. The Schwarzschild region is on the left-hand side of the
  boundary, and the FLRW region in on the right-hand side. It is
  assumed that the cosmological region possesses flat spatial
  sections, so that the boundary expands forever. The portion of the
  spacetime bounded by the past singularity, the future horizon, and
  future null infinity can be covered by the $u$ and $v$ coordinates
  introduced in the text. In the diagram, $u$ increases at 45 degrees
  toward the left, while $v$ increases at 45 degrees toward the
  right. The future horizon is located at $u=\infty$, future null
  infinity at $v=\infty$, the boundary at $v - u = 2\chi_0$, and the
  past horizon at $v = 3\chi_0$.}  
\end{figure} 

\subsection{Joining the metrics at the boundary} 

The metric inside $\Sigma$ is expressed in the usual Schwarzschild
coordinates as 
\begin{equation}
ds_{\rm in}^2 = -f\, dt^2 + f^{-1}\, dr^2 + r^2\, d\Omega^2,  
\label{2.1}
\end{equation}
where $f = 1 - 2M/r$ and $d\Omega^2 = d\theta^2 + \sin^2\theta\,
d\phi^2$; $M$ is the gravitational mass of the black hole. The metric
of Eq.~(\ref{2.1}) is valid for $r < r_\Sigma(\tau)$, where $r_\Sigma$
is the changing radius of the timelike hypersurface $\Sigma$; $\tau$
is proper time for observers comoving with the boundary. Because the
hypersurface is generated by timelike geodesics, its equations of
motion are $\dot{t} = \tilde{E}/f$ and $\dot{r}^2 + f =
\tilde{E}^2$. Here, an overdot indicates differentiation with
respect to $\tau$, and $\tilde{E}$ is the usual conserved-energy
parameter. 

For bound motion ($\tilde{E} < 1$), the solution to the
equations of motion can be expressed in the parametric form  
\begin{eqnarray}
r_\Sigma(\eta) &=& \frac{1}{2}\, r_{\rm max} (1-\cos\eta), 
\nonumber \\ 
& & \label{2.2} \\ 
\tau(\eta) &=& \frac{1}{2}\, r_{\rm max} \sqrt{r_{\rm max}/2M}
(\eta - \sin\eta),  
\nonumber
\end{eqnarray} 
in terms of a conformal-time parameter $\eta$; the constant 
$r_{\rm max} \equiv 2M/(1-\tilde{E}^2)$ represents the boundary's
maximum radius. For unbound motion ($\tilde{E} > 1$), the solution is  
\begin{eqnarray}
r_\Sigma(\eta) &=& \frac{1}{2}\, r_0 (\cosh\eta-1), 
\nonumber \\
& & \label{2.3} \\
\tau(\eta) &=& \frac{1}{2}\, r_0 \sqrt{r_0/2M} (\sinh\eta-\eta), 
\nonumber 
\end{eqnarray} 
where $r_0 \equiv 2M/(\tilde{E}^2-1)$. Finally, for marginally bound
motion ($\tilde{E}=1$) we have
\begin{equation} 
r_\Sigma(\tau) = \frac{9M}{2}\, \biggl( \frac{2\tau}{9M}
\biggr)^{2/3}. 
\label{2.4}
\end{equation} 

The metric outside $\Sigma$ is expressed as 
\begin{equation}
ds_{\rm out}^2 = a^2(\eta) \bigl[ -d\eta^2 + d\chi^2 + s^2(\chi)\,
d\Omega^2 \bigr], 
\label{2.5}
\end{equation} 
where $\eta$ is conformal time, $\chi$ a radial coordinate, $a(\eta)$
the cosmological scale factor, and $s(\chi)$ a function that
determines the geometry of the spatial sections: 
\begin{equation}
s(\chi) = \left\{ 
\begin{array}{cl} 
\sin\chi  & \qquad k = 1 \\
\chi      & \qquad k = 0 \\
\sinh\chi & \qquad k = -1  
\end{array}
\right. .
\label{2.6}
\end{equation} 
The metric of Eq.~(\ref{2.5}) is valid for $\chi > \chi_0$, with
$\chi_0$ denoting the comoving position of $\Sigma$. The behavior of
the scale factor is determined by the spatial curvature and the matter
content of the universe. Here we assume that the universe is matter
dominated, with a stress-energy tensor given by $T^{\alpha\beta} =
\rho u^\alpha u^\beta$, where $\rho$ is the dust's mass density and
$u^\alpha = a^{-1} \partial x^\alpha/\partial\eta$ its
four-velocity. Conservation of energy implies that 
$\rho a^3 = \mbox{constant} \equiv 3C/(8\pi)$. By
virtue of the Einstein field equations, the scale factor must satisfy
$a^{\prime 2} + k a^2 = C a$, in which a prime indicates
differentiation with respect to $\eta$. Proper time for comoving
observers is obtained by integrating $d\tau = a(\eta)\, d\eta$. 

For spatial sections with positive curvature ($k=1$), the scale factor
and the proper time are given by 
\begin{equation}
a(\eta) = \frac{C}{2}\, (1 - \cos\eta), \qquad
\tau(\eta) = \frac{C}{2}\, (\eta - \sin\eta), 
\label{2.7}
\end{equation}
where $C = 8\pi \rho a^3/3$ is a constant. For spatial sections with
negative curvature ($k=-1$), we have instead 
\begin{equation}
a(\eta) = \frac{C}{2}\, (\cosh\eta - 1), \qquad
\tau(\eta) = \frac{C}{2}\, (\sinh\eta - \eta). 
\label{2.8}
\end{equation}
Finally, for flat spatial sections ($k=0$) we have 
\begin{equation}
a(\eta) = \frac{C}{4}\, \eta^2, \qquad
\tau(\eta) = \frac{C}{12}\, \eta^3. 
\label{2.9}
\end{equation} 

The metrics of Eq.~(\ref{2.1}) and (\ref{2.2}) must be matched at
their common boundary \cite{23}, the timelike hypersurface $\Sigma$
described by $r = r_\Sigma(\tau)$ on the Schwarzschild side, and $\chi
= \chi_0$ on the FLRW side. The boundary's induced metric, as seen
from the interior, is 
\begin{equation}
ds^2_\Sigma = -d\tau^2 + \bigl[r_\Sigma(\tau)\bigr]^2\, d\Omega^2. 
\label{2.10}
\end{equation}
As seen from the exterior, it is 
\begin{equation}
ds^2_\Sigma = -d\tau^2 + \bigl[a(\tau) s(\chi_0) \bigr]^2\, d\Omega^2.  
\label{2.11}
\end{equation}
In both cases the induced metric is expressed in terms of the
hypersurface's intrinsic coordinates $(\tau,\theta,\phi)$, with $\tau$
denoting proper time for observers comoving with the
hypersurface. Equality of the induced metric on both sides of the
hypersurface implies
\begin{equation}
r_\Sigma(\tau) = a(\tau) s(\chi_0),  
\label{2.12}
\end{equation}
in which $r_\Sigma(\tau)$ is given implicitly by Eqs.~(\ref{2.2}),
(\ref{2.3}), or (\ref{2.4}), while $a(\tau)$ is given implicitly by
Eqs.~(\ref{2.7}), (\ref{2.8}), or (\ref{2.9}). These equations are 
compatible with Eq.~(\ref{2.12}) provided that the relations 
\begin{equation}
\tilde{E} = \frac{ds}{d\chi}(\chi_0)
\label{2.13}
\end{equation} 
and
\begin{equation}
M = \frac{1}{2}\, C s^3(\chi_0) = 
\frac{4\pi}{3}\, \rho a^3 s^3(\chi_0)   
\label{2.14}
\end{equation} 
hold. Equation (\ref{2.14}) implies that the mass of the black hole is
equal to the mass removed from the interior of $\Sigma$, as was said
previously. It can be checked that Eqs.~(\ref{2.12}) and (\ref{2.13})
ensure the continuity of the extrinsic curvature across $\Sigma$: the
union of the Schwarzschild and FLRW metrics forms a globally valid
solution to the Einstein field equations \cite{23}.      

\subsection{Double-null coordinates} 

The metric of the Einstein-Straus spacetime has thus far been
expressed in terms of two coordinates patches. We now construct a
single coordinate system $(u,v,\theta,\phi)$ that covers both regions
of the spacetime. These coordinates are null, and in the cosmological
region they are defined by 
\begin{equation}
u = \eta - \chi, \qquad
v = \eta + \chi. 
\label{2.15}
\end{equation} 
They are related to similar coordinates that can be introduced in the
Schwarzschild region, 
\begin{equation}
\bar{u} = t - r^*, \qquad
\bar{v} = t + r^*,
\label{2.16}
\end{equation}
where $r^* = \int f^{-1}\, dr = r + 2M\ln(r/2M - 1)$. The relations
$\bar{u}(u)$, $\bar{v}(v)$ can be obtained by using these coordinates
to express continuity of the induced metric across $\Sigma$.  

For example, equating the $ds_\Sigma^2$ of Eq.~(\ref{2.11}), expressed
in terms of $d\bar{v}$, to the $ds_\Sigma^2$ of Eq.~(\ref{2.10}),
expressed in terms of $dv$, yields
\begin{equation}
(1-2M/r_\Sigma) d\bar{v}^2 - 2 dr_\Sigma d\bar{v} = a^2 dv^2.  
\label{2.17}
\end{equation}
Here, the scale factor must be expressed as a function of $v -
\chi_0$, and $r_\Sigma$ is viewed as the function of $v$ defined by  
Eq.~(\ref{2.12}). Equation (\ref{2.17}) is a quadratic equation for
$d\bar{v}/dv$. Solving this, using Eq.~(\ref{2.14}), gives 
\begin{equation}
\frac{d\bar{v}}{dv} = \frac{a' s_0 + a \sqrt{1-k {s_0}^2}}{1 - C
  {s_0}^2/a}, 
\label{2.18}
\end{equation} 
where a prime indicates differentiation with respect to $v-\chi_0$,
$s_0 \equiv s(\chi_0)$, and $C = 8\pi \rho a^3/3$. More explicitly,
substituting Eqs.~(\ref{2.6}), (\ref{2.7}), and (\ref{2.14}) into
Eq.~(\ref{2.18}) yields  
\begin{equation} 
\frac{d\bar{v}}{dv} = \frac{M}{\sin^3\chi_0}\, \frac{
  \bigl[1-\cos(v-\chi_0)\bigr] \bigl[\cos\chi_0 - \cos v]} 
  { \cos(2\chi_0) - \cos(v-\chi_0) }
\label{2.19}
\end{equation}
for $k = 1$. Similarly, substituting Eqs.~(\ref{2.6}), (\ref{2.8}),
and (\ref{2.14}) into Eq.~(\ref{2.18}) yields 
\begin{equation} 
\frac{d\bar{v}}{dv} = \frac{M}{\sinh^3\chi_0}\, \frac{
  \bigl[\cosh(v-\chi_0) - 1\bigr] \bigl[\cosh v - \cosh \chi_0]}  
  { \cosh(v-\chi_0) - \cosh(2\chi_0) } 
\label{2.20}
\end{equation}
for $k = -1$. Finally, using Eq.~(\ref{2.9}) instead gives 
\begin{equation} 
\frac{d\bar{v}}{dv} = \frac{M}{2 {\chi_0}^3}\, \frac{(v-\chi_0)^3}
  {v - 3 \chi_0}
\label{2.21}
\end{equation}
for $k = 0$. These relations for $d\bar{v}/dv$ go to zero at   
$v = \chi_0$, which represents the past singularity of the
cosmological spacetime, and they are singular at $v = 3\chi_0$, which
represents the extension of the black hole's past horizon into the
cosmological region (see Fig.~1).  

A nearly identical set of calculations gives us the relations
$\bar{u}(u)$. Here we find  
\begin{equation} 
\frac{d\bar{u}}{du} = \frac{M}{\sin^3\chi_0}\, \frac{
  \bigl[1-\cos(u+\chi_0)\bigr] \bigl[\cos\chi_0 - \cos u]} 
  { \cos(2\chi_0) - \cos(u+\chi_0) }
\label{2.22}
\end{equation}
for $k = 1$, 
\begin{equation} 
\frac{d\bar{u}}{du} = \frac{M}{\sinh^3\chi_0}\, \frac{
  \bigl[\cosh(u+\chi_0) - 1\bigr] \bigl[\cosh u - \cosh \chi_0]}  
  { \cosh(u+\chi_0) - \cosh(2\chi_0) } 
\label{2.23}
\end{equation}
for $k = -1$, and 
\begin{equation} 
\frac{d\bar{u}}{du} = \frac{M}{2 {\chi_0}^3}\, \frac{(u+\chi_0)^3}
 {u + 3 \chi_0}
\label{2.24}
\end{equation}
for $k = 0$.
  
Equations (\ref{2.19})--(\ref{2.24}) can all be integrated in closed
form. For $k = \pm 1$, the resulting expressions for $\bar{v}(v)$ and
$\bar{u}(u)$ are quite complicated, and we shall not display them
here. For $k = 0$ the relations are simple, and we find 
\begin{eqnarray}
\bar{v} &=& M \Biggl[ \frac{1}{6}
\biggl(\frac{v}{\chi_0}\biggr)^3 
+ \frac{3}{2} \biggl(\frac{v}{\chi_0}\biggr) 
+ 4 \ln \biggl( \frac{v}{\chi_0} - 3 \biggr) 
\nonumber \\ & & \mbox{} 
- \biggl(\frac{10}{3}  + 8\ln 2 \biggr) \Biggr]  
\label{2.25}
\end{eqnarray}
and 
\begin{equation}
\bar{u} = M \Biggl[ \frac{1}{6}
\biggl(\frac{u}{\chi_0}\biggr)^3 
+ \frac{3}{2} \biggl(\frac{u}{\chi_0}\biggr) 
- 4 \ln \biggl( \frac{u}{\chi_0} + 3 \biggr) \Biggr]. 
\label{2.26}
\end{equation}
We have tuned the constants of integration to produce agreement
between Eqs.~(\ref{2.25}) and (\ref{2.26}) and the defining relation
$\frac{1}{2}(\bar{v}-\bar{u}) = r + 2M\ln(r/2M - 1)$; the 
right-hand side must be evaluated on $\Sigma$ and expressed in terms
of $u$ and $v$ by means of Eq.~(\ref{2.12}). The relation $\bar{u}(u)$
is plotted in Fig.~2; the relation between $\bar{v}$ and $v$ is nearly
identical.  

\begin{figure}
\special{hscale=35 vscale=35 hoffset=-20.0 voffset=10.0
         angle=-90.0 psfile=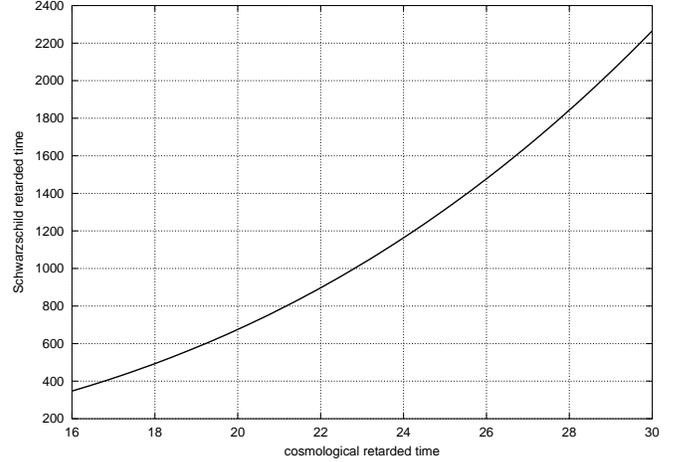}
\vspace*{2.6in}
\caption{The relationship between the cosmological retarded-time
  coordinate $u$ and the Schwarzschild retarded-time coordinate
  $\bar{u}$. The plot is constructed for the interval $16 < u < 30$,
  choosing $\chi_0 = 1$ and setting $2M \equiv 1$. A graph of
  $\bar{v}(v)$ would be almost identical for these choices of 
  parameters.} 
\end{figure} 

Both sides of the Einstein-Straus spacetime can be covered by the
single coordinate system $(u,v,\theta,\phi)$ constructed above. Our 
coordinates, however, do no extend beyond the future horizon of the
Schwarzschild region. As far as we are aware, this coordinate system
was never presented before in the literature. 
  
\section{Scalar wave equation} 

We consider the propagation of a scalar field $\Phi$ in the
Einstein-Straus spacetime. This field satisfies the wave equation  
\begin{equation}
(\Box - \xi R) \Phi = 0, 
\label{3.1}
\end{equation} 
in which $\Box = g^{\alpha\beta} \nabla_\alpha \nabla_\beta$ is the
curved-spacetime d'Alembertian operator, $R$ the Ricci scalar, and
$\xi$ an arbitrary dimensionless constant; the scalar field is
minimally coupled to curvature if $\xi = 0$, and it is conformally
coupled if $\xi = \frac{1}{6}$. 

In the cosmological region, where the metric is given by
Eq.~(\ref{2.5}), we separate the variables by writing 
\begin{equation} 
\Phi(\eta,\chi,\theta,\phi) = \frac{1}{a(\eta) s(\chi)}\, 
\sum_{lm} \psi_l(\eta,\chi)\, Y_{lm}(\theta,\phi).   
\label{3.2}
\end{equation}
Here, $Y_{lm}(\theta,\phi)$ are the usual spherical harmonics, and
$\psi_l(\eta,\chi)$ is a reduced wave function that satisfies 
\begin{equation}
\biggl[ -\frac{\partial^2}{\partial \eta^2} 
+ \frac{\partial^2}{\partial \chi^2} + \frac{1-6\xi}{2a(\eta)/C} 
- \frac{l(l+1)}{s^2(\chi)} \biggr] \psi_l = 0,  
\label{3.3}
\end{equation} 
where $2a(\eta)/C$ and $s(\chi)$ are respectively equal to
$1-\cos\eta$ and $\sin\chi$ if $k=-1$, $\cosh\eta - 1$ and $\sinh\chi$
if $k = 1$, and $\frac{1}{2} \eta^2$ and $\chi$ if $k = 0$.     

In the Schwarzschild region, where the metric is given by
Eq.~(\ref{2.1}), we let 
\begin{equation}
\Phi(t,r,\theta,\phi) = \frac{1}{r}\, \sum_{lm} \psi_l(t,r)\,
Y_{lm}(\theta,\phi). 
\label{3.4}
\end{equation}
The reduced wave function now satisfies 
\begin{equation}
\biggl\{ -\frac{\partial^2}{\partial t^2} 
+ \frac{\partial^2}{\partial r^{* 2}} - 
f \biggl[ \frac{l(l+1)}{r^2} + \frac{2M}{r^3} \biggr] 
 \biggr\} \psi_l = 0, 
\label{3.5}
\end{equation}
where $r^* = r + 2M\ln(r/2M - 1)$ and $f = 1 - 2M/r$.  

The field $\Phi$ and its derivatives $\Phi_{,\alpha}$ are continuous
across the boundary separating the Schwarzschild and cosmological
regions. Smoothness of the metric across $\Sigma$ ensures that the
reduced wave function $\psi_l$ and its derivatives are also continuous
at the boundary. 

Using the coordinates $u$ and $v$ introduced in Sec.~II B, the reduced
wave equation becomes   
\begin{equation}
\frac{\partial^2 \psi_l}{\partial u \partial v} = 
-\frac{1}{4} V_l(u,v)\, \psi_l 
\label{3.6}
\end{equation}
on both sides of the boundary. The effective potential $V_l(u,v)$
takes a different form in the two portions of the spacetime. In the
cosmological region, it is 
\begin{equation}
V^{\rm out}_l = \frac{l(l+1)}{s^2(\chi)} - \frac{1-6\xi}{2a(\eta)/C},  
\label{3.7}
\end{equation}
where $\eta = \frac{1}{2}(v+u)$ and $\chi = \frac{1}{2}(v-u)$. In the
Schwarzschild region, we have instead 
\begin{equation}
V^{\rm in}_l = \frac{d \bar{u}}{du} \frac{d \bar{v}}{dv}\, 
\biggl( 1 - \frac{2M}{r} \biggr) 
\biggl[ \frac{l(l+1)}{r^2} + \frac{2M}{r^3} \biggr], 
\label{3.8}
\end{equation}
in which $r(u,v)$ is defined implicitly by $r^* = \frac{1}{2}(\bar{v} 
- \bar{u})$; the relations $\bar{u}(u)$ and $\bar{v}(v)$ were obtained
at the end of Sec.~II B. It should be noted that the potential is
{\it discontinuous} at the boundary; this is explained by the fact
that the density $\rho$ is discontinuous at $\Sigma$.   

The reduced wave equation, in the form of Eq.~(\ref{3.6}), can be
straightforwardly integrated on both sides of $\Sigma$, which is
described by the relation $v = u + 2\chi_0$. The initial data
required for a unique solution consists of $\psi_l(u,v=v_0)$ and
$\psi_l(u=u_0,v)$, the value of the wave function on the two null
surfaces $v = v_0$ and $u = u_0$; these values can be specified
freely.    

\section{Numerical results} 

We now present our numerical integrations of Eq.~(\ref{3.6}). The
algorithm used for this task is based on the finite-difference scheme
suggested by Gundlach, Price, and Pullin \cite{11}. In this scheme,
the $(u,v)$ continuum is discretized in units of $\Delta$, and
Eq.~(\ref{3.6}) is replaced by the approximation   
\begin{eqnarray}
\psi_l(u+\Delta,v+\Delta) &=& \biggl[ 1 - \frac{1}{8} V_l\, \Delta^2  
+ O(\Delta^4) \biggr] \Bigl[ \psi_l(u+\Delta,v) \nonumber \\
& & \mbox{} + \psi_l(u,v+\Delta) \Bigr] 
- \psi_l(u,v), 
\label{4.1}
\end{eqnarray} 
in which the potential is evaluated at the off-grid point $(u +
\frac{1}{2}\Delta, v + \frac{1}{2} \Delta)$. For simplicity we
restrict our attention to spatially-flat cosmologies ($k = 0$), and a
scalar field that is minimally coupled to curvature ($\xi = 0$). In
Eq.~(\ref{3.7}) we must therefore substitute $s(\chi) = \chi =
\frac{1}{2}(v-u)$ and $2a(\eta)/C = \frac{1}{2} \eta^2 = \frac{1}{8}
(v+u)^2$. In Eq.~(\ref{3.8}), the relations $\bar{v}(v)$ and
$\bar{u}(u)$ are obtained from Eq.~(\ref{2.25}) and (\ref{2.26}), 
respectively; their derivatives are given by Eqs.~(\ref{2.21}) and  
(\ref{2.24}). For the initial value of the wave function on $u = u_0$
we choose a gaussian wave packet of width $\sigma$ centered at $v =
v_c$:  
\begin{equation}
\psi_l(u=u_0,v) = \exp\biggl[ -\frac{(v-v_c)^2}{2\sigma^2} \biggr].  
\label{4.2}
\end{equation} 
On $v = v_0$ we make the choice $\psi_l(u,v=v_0) = \psi_l(u_0,v_0)
\simeq 0$: the initial field is a purely ingoing wave packet.  

The parameter space associated with the numerical integration of
Eq.~(\ref{3.6}) is rather large, but moving in this space does not
affect the qualitative aspects of the wave's evolution. For
concreteness we set $\chi_0 = 1$, $2M = 1$, $u_0 = 17$, $v_0 = 16$,
$v_c = 16.9$, $\sigma = 0.2$. These choices ensure that the initial
wave packet is contained entirely within the Schwarzschild region of
the spacetime; its center is located in the strong-field region near
$r^* = 0$. The discretization unit $\Delta$ must be set sufficiently
small to ensure that the potential $V_l$ in Eq.~(\ref{4.1}) is well
sampled. In our simulations we use $\Delta = 0.001$; choosing a
smaller value does not appreciably change our results. In Fig.~3 we
display the numerical grid and a few relevant features of the
spacetime.    

\begin{figure}
\special{hscale=35 vscale=35 hoffset=-20.0 voffset=10.0
         angle=-90.0 psfile=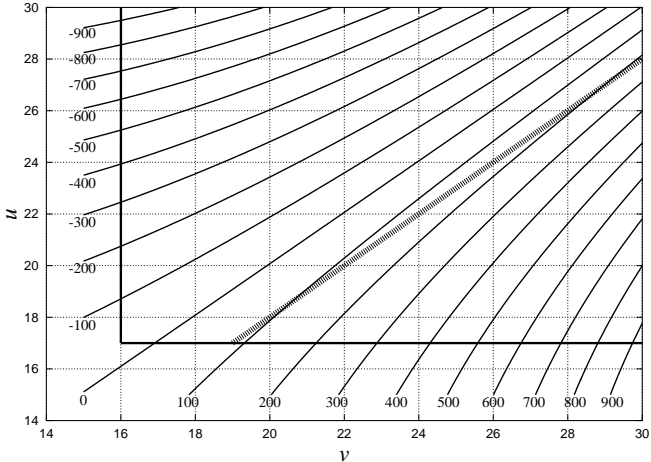}
\vspace*{2.6in}
\caption{The portion of the Einstein-Straus spacetime that is relevant
         for our numerical integrations. With $\chi_0 = 1$, this
         corresponds to the intervals $17 < u < 30$ and $16 < v <
         30$. The thick, dashed line described by $v = u + 2$
         represents the boundary $\Sigma$. The 
         Schwarzschild region of the spacetime appears above the
         boundary, while the FLRW region appears below. The solid
         lines are curves $r^*(u,v) = \mbox{constant}$, with the 
         label giving the value of the constant (in units $2M = 1$); 
         these curves are meaningful only in the Schwarzschild part of
         the spacetime. Notice that the curve $r^* = 0$ is well
         approximated by $v = u$. The initial wave packet on $u = 17$
         is centered at $v = 16.9$, at which $r^* \simeq 0$; it is
         therefore located in the strong-field region of the
         Schwarzschild spacetime. Because $r^*$ is negatively large
         everywhere on $u = 30$, this surface closely approximates the
         black hole's event horizon. We similarly take $v = 30$ to
         approximate future null infinity, although the approximation
         becomes poor as $u$ increases.}    
\end{figure} 

Equation (\ref{4.1}) is used to numerically evolve $\psi_l$ from the
initial conditions imposed on the lines $u = 17$ and $v = 16$. The
wave function is then evaluated on three different curves of the
$(u,v)$ plane. The first is the line $u = 30$, which approximates the
black hole's event horizon (see Fig.~3); there $\psi_l$ is expressed
as a function of cosmological advanced time $v$. The second is the
line $v = u + 2$, which represents the boundary $\Sigma$; there
$\psi_l$ is expressed as a function of conformal time $\eta = v - 1 =
u + 1$. The third is the line $v = 30$, which approximates future null
infinity (see Fig.~3); there $\psi_l$ is expressed as a function of
cosmological retarded time $u$. The results of our numerical
simulations are displayed in Figs.~4--12, and the captions describe
the graphs in detail. In the following paragraphs we describe and
explain the salient features. 

\begin{figure}
\special{hscale=35 vscale=35 hoffset=-20.0 voffset=10.0
         angle=-90.0 psfile=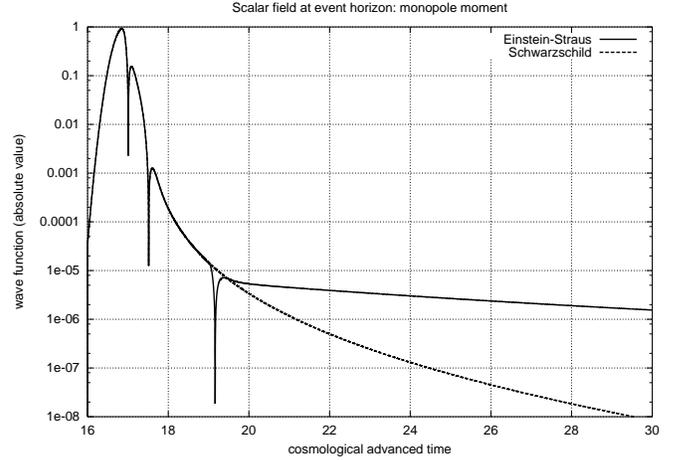}
\vspace*{2.6in}
\caption{A plot of $|\psi_0(u=30)|$, the absolute value of the 
$l = 0$ wave function on the event horizon, as a function of
cosmological advanced time $v$. Prior to $v = 19$, the wave behaves as
it would in Schwarzschild spacetime. This early-time behavior is
characterized by quasi-normal oscillations (which on a logarithmic
scale are revealed by deep troughs) and the onset of inverse power-law
falloff. At $v = 19$ the field comes in causal contact with the
cosmological region and suddenly alters its behavior. The wave
function changes sign, and it subsequently falls off at a much slower
rate; as we show in Sec.~V, this late-time behavior is not well 
described by an inverse power-law.}    
\end{figure} 

\begin{figure}
\special{hscale=35 vscale=35 hoffset=-20.0 voffset=10.0
         angle=-90.0 psfile=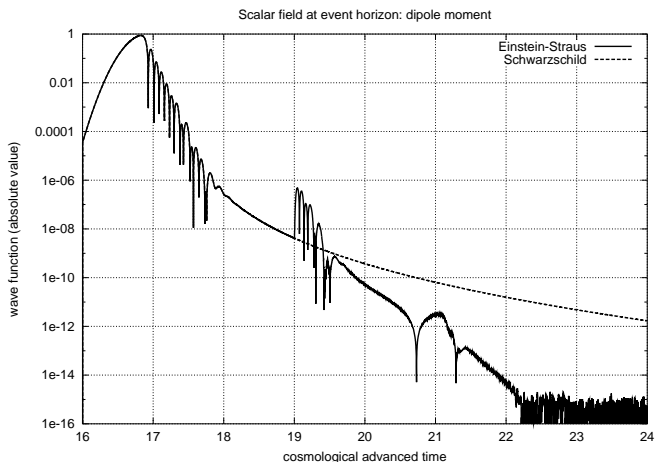}
\vspace*{2.6in}
\caption{A plot of $|\psi_1(u=30)|$, the absolute value of the 
$l = 1$ wave function on the event horizon, as a function of
cosmological advanced time $v$. The early-time behavior is again
characterized by quasi-normal oscillations (with a higher frequency
and a faster decay time than the $l = 0$ case) and the onset of
inverse power-law falloff. At $v = 19$ a faint echo of the early-time
signal registers on the event horizon. A second echo also appears  
near $v = 21$. Beyond $v=22$, the wave function drops below $10^{-15}$
and our numerical results become unreliable.}  
\end{figure} 

\begin{figure}
\special{hscale=35 vscale=35 hoffset=-20.0 voffset=10.0
         angle=-90.0 psfile=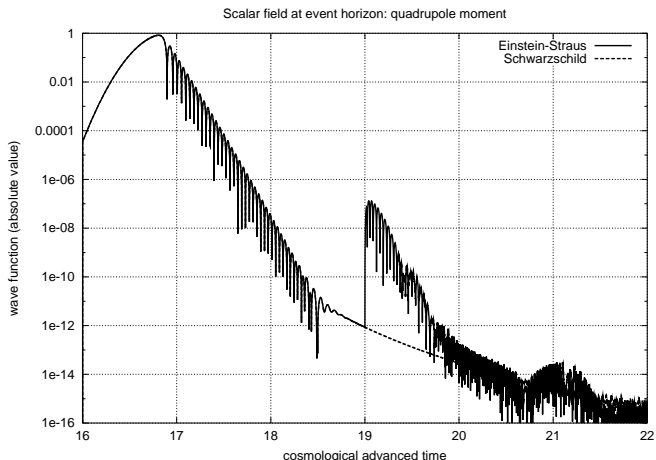}
\vspace*{2.6in}
\caption{A plot of $|\psi_2(u=30)|$, the absolute value of the 
$l = 2$ wave function on the event horizon, as a function of
cosmological advanced time $v$. Again we have quasi-normal
oscillations and inverse power-law falloff at early times, 
followed by echos of the early signal at later times. Our numerical
results become noisy beyond $v=20$, but a second echo can still be
seen near $v=21$.}        
\end{figure} 

\begin{figure}
\special{hscale=35 vscale=35 hoffset=-20.0 voffset=10.0
         angle=-90.0 psfile=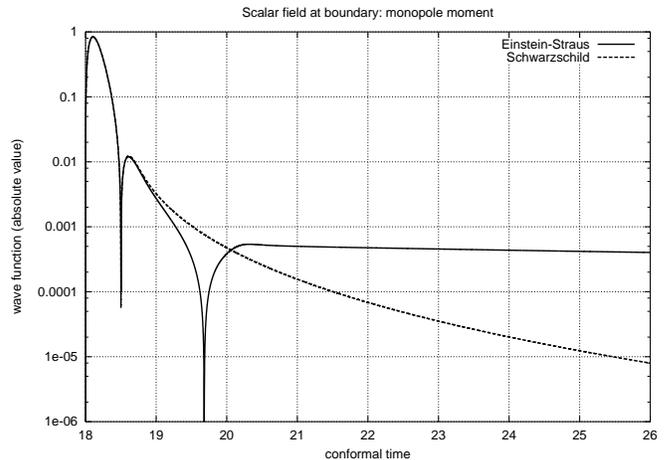}
\vspace*{2.6in}
\caption{A plot of $|\psi_0(v = u + 2)|$, the absolute value of the  
$l = 0$ wave function on the boundary $\Sigma$, as a function of
conformal time $\eta$. At early times the field behaves as it would 
in Schwarzschild spacetime. At later times, from $\eta = 19$
onward, the multiple reflections from the potential barrier near $r^*
= 0$ and the jump discontinuity at $\Sigma$ alter the behavior of the
wave function. At $\eta \simeq 19.7$, the wave function changes sign,
and it subsequently falls off at a slow rate; as we show in Sec.~V,
this behavior is not well described by an inverse power-law.}      
\end{figure} 

\begin{figure}
\special{hscale=35 vscale=35 hoffset=-20.0 voffset=10.0
         angle=-90.0 psfile=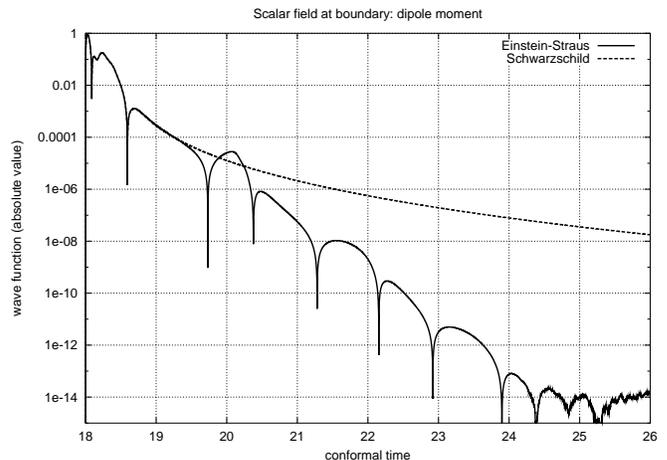}
\vspace*{2.6in}
\caption{A plot of $|\psi_1(v = u + 2)|$, the absolute value of the  
$l = 1$ wave function on the boundary $\Sigma$, as a function of
conformal time $\eta$. Again we have Schwarzschild-type behavior at
early times, and strong deviations from it starting at $\eta = 19$. 
The multiple reflections of the wave packet give rise to the
oscillatory behavior seen in the figure. The period of these 
oscillations is approximately $\Delta \eta \simeq 2$, twice the
light-travel time between the boundary and the potential barrier. (At
first glance, the frequency appears to be twice as large because the
repeated pattern, which can be seen in the interval $18 < \eta < 19$,
already contains a sign change.) The end result is a field that decays
faster than in Schwarzschild spacetime, contrary to what takes place
for $l = 0$. Our numerical results become unreliable beyond $\eta =
24$.}     
\end{figure} 

\begin{figure}
\special{hscale=35 vscale=35 hoffset=-20.0 voffset=10.0
         angle=-90.0 psfile=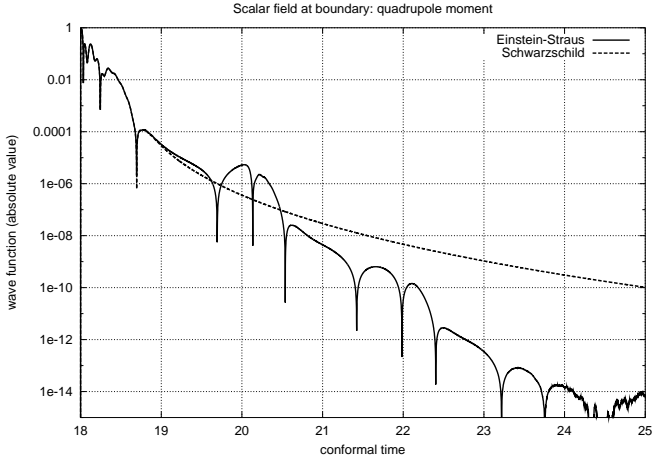}
\vspace*{2.6in}
\caption{A plot of $|\psi_2(v = u + 2)|$, the absolute value of the  
$l = 2$ wave function on the boundary $\Sigma$, as a function of
conformal time $\eta$. We recognize the late-time oscillatory behavior 
that was identified in Fig.~8. Here also the field decays faster than
it would in Schwarzschild spacetime. Our numerical results become
unreliable beyond $\eta = 24$.}      
\end{figure} 

\begin{figure}
\special{hscale=35 vscale=35 hoffset=-20.0 voffset=10.0
         angle=-90.0 psfile=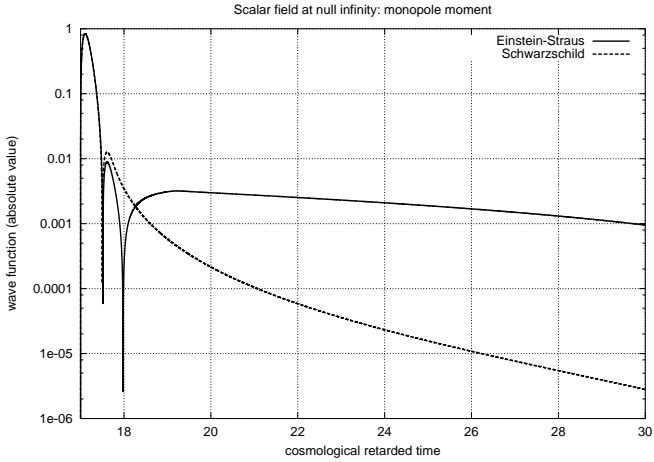}
\vspace*{2.6in}
\caption{A plot of $|\psi_0(v = 30)|$, the absolute value of the  
$l = 0$ wave function at future null infinity, as a function of
cosmological retarded time $u$. The behavior seen here is more or less
a direct image of what was already seen at the boundary, in
Fig.~7. Compared with the Schwarzschild behavior, the field decays at
a slower rate; as we show in Sec.~V, this behavior is not well
described by an inverse power-law.}       
\end{figure} 

\begin{figure}
\special{hscale=35 vscale=35 hoffset=-20.0 voffset=10.0
         angle=-90.0 psfile=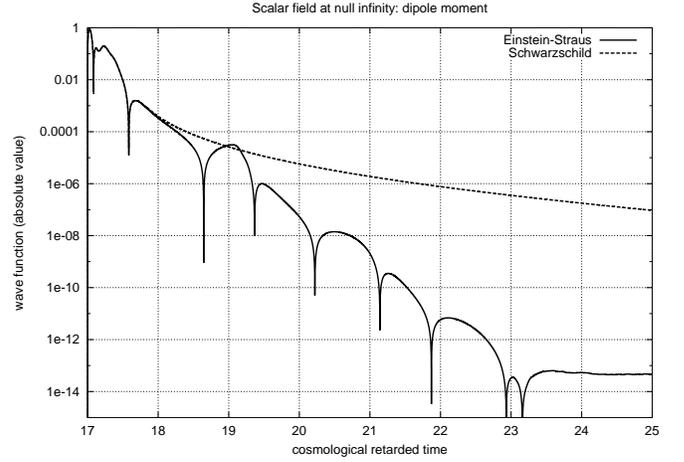}
\vspace*{2.6in}
\caption{A plot of $|\psi_1(v = 30)|$, the absolute value of the  
$l = 1$ wave function at future null infinity, as a function of
cosmological retarded time $u$. The behavior seen here is more or less
a direct image of what was already seen at the boundary, in
Fig.~8. Contrary to the $l = 0$ case, the field decays faster in the
Einstein-Straus spacetime than in Schwarzschild spacetime.}          
\end{figure} 

\begin{figure}
\special{hscale=35 vscale=35 hoffset=-20.0 voffset=10.0
         angle=-90.0 psfile=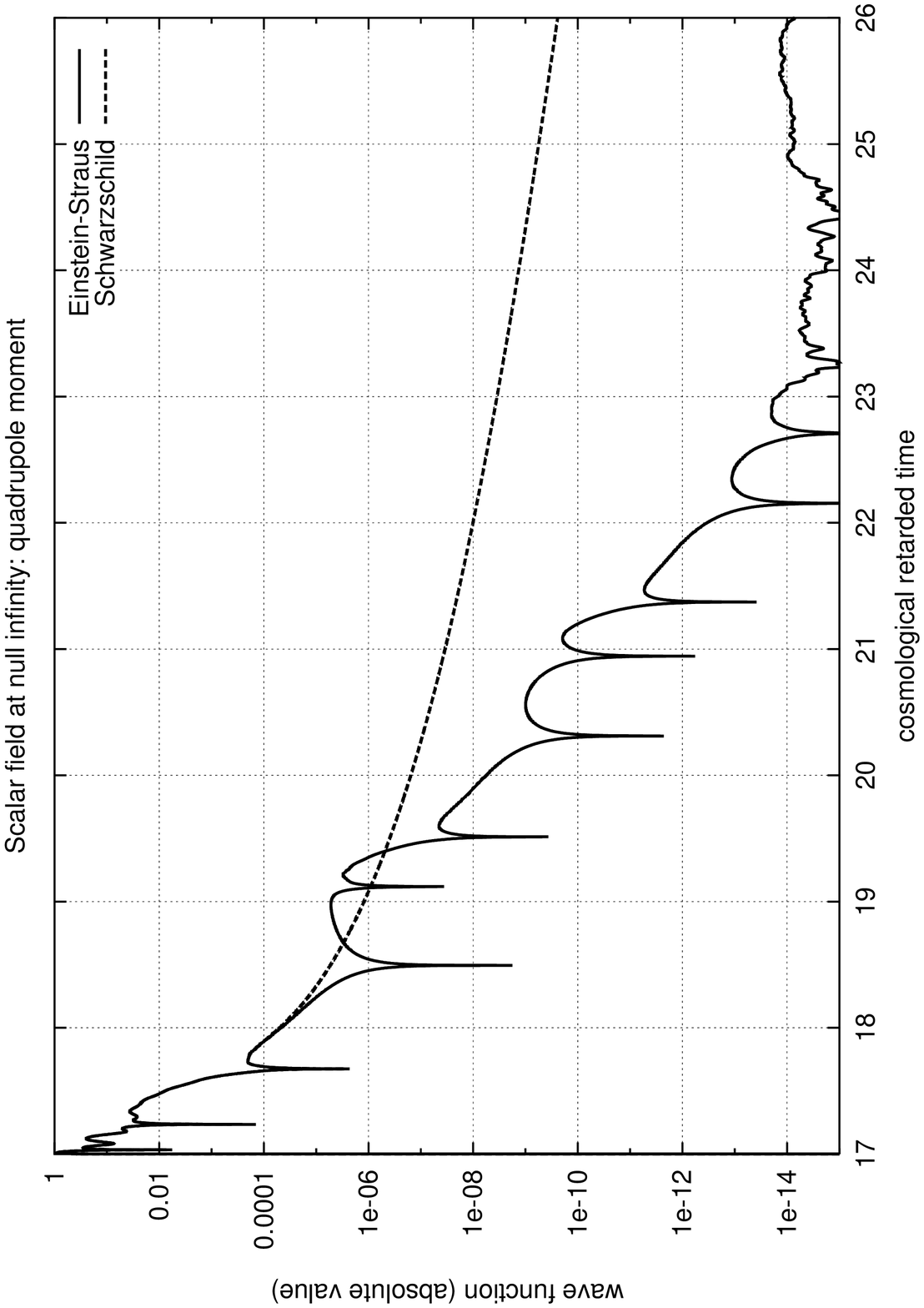}
\vspace*{2.6in}
\caption{A plot of $|\psi_2(v = 30)|$, the absolute value of the  
$l = 2$ wave function at future null infinity, as a function of
cosmological retarded time $u$. The behavior seen here is more or less
a direct image of what was already seen at the boundary, in
Fig.~9. Compared with its Schwarzschild behavior, the field decays at
a faster rate.}         
\end{figure} 

Wave propagation in the Einstein-Straus spacetime is governed by the
potential $V_l$ displayed in Eqs.~(\ref{3.7}) and (\ref{3.8}). The
most important aspects of this potential are that it peaks near 
$r^* = 0$ (because of the strong curvature near the black hole) and
that it is discontinuous at the boundary (because of the jump in the
density $\rho$). For the purposes of understanding the gross
properties of the wave's evolution, the potential can be thought of as
consisting of a narrow barrier near $r^* = 0$ and a jump discontinuity
at $\chi = \chi_0$. After each encounter with the barrier or the jump,
the wave is partially transmitted and reflected, and these processes 
determine the late-time behavior of the scalar field. The diagram of
Fig.~3 informs us that the curve $r^* = 0$ is well approximated by the
line $v = u$. The potential barrier is therefore located near $v = u$, 
while the jump discontinuity is at $v = u + 2$.    

Consider, as we do in our numerical simulations, an initially ingoing
wave packet on $u = 17$, centered at $v = 16.9$. In the absence of any
potential, the wave would keep its shape as it moves toward increasing
values of $u$. In the diagram of Fig.~3, the wave packet would move
vertically upward, and it would eventually register on the ``event
horizon'' at $u = 30$; the wave profile on $u = 30$ would be identical
to the initial configuration on $u = 17$. Instead, our wave packet
moves in the presence of the potential, which produces partial
transmissions and reflections as well as distortions in the
packet's shape. 

Because the initial wave packet is centered near $r^* = 0$, it
almost immediately encounters the potential barrier as it propagates
away from $u = 17$. The partially transmitted wave moves vertically
upward in the diagram of Fig.~3, and a distorted wave profile
registers on the event horizon; this profile still has a recognizable
center at $v = 16.9$. On the other hand, the partially reflected wave
moves horizontally toward the right, and it soon arrives at the
boundary; the center of the wave packet arrives shortly after $\eta =
u_0 + 1 = 18$. 

At the boundary, the right-moving wave encounters the jump in the
potential and undergoes partial transmission and reflection. The
transmitted wave proceeds toward future null infinity without much
further distortion; the center of the wave packet arrives shortly
after $u = u_0 = 17$. On the other hand, the reflected wave moves
vertically upward in the diagram, and the center of the packet follows 
the line $v = \eta_0 + 1 = 19$. It encounters once more the potential
barrier near $u = v$. 

At the barrier, the up-moving wave is partially transmitted and
reflected. The transmitted wave continues along $v = 19$ and it
finally registers on the event horizon, where it is recognized as a
faint echo of the original signal. The reflected wave moves
horizontally along $u = 19$, and proceeds toward the jump at
$v = u + 2$. There the entire cycle of partial transmissions and
reflections repeats once more.  

In the figures, the behavior of the wave function in the
Einstein-Straus spacetime is contrasted with the behavior it would
have in Schwarzschild spacetime. Because the scalar field is at
first causally disconnected from the cosmological region, the
early-time behaviors coincide. As the field comes in causal
contact with the boundary and the expanding universe beyond it, its 
behavior changes. In particular, the inverse power-law falloff
witnessed at late times for Schwarzschild spacetime is not seen in the
Einstein-Straus spacetime. Instead, the field's late-time behavior is
quite complicated, and governed by many cycles of partial
transmissions and reflections, as was described above. 

\section{Partial transmissions and reflections: A toy model}  

In this final section we firm up the claim made in Sec.~IV that
late-time wave propagation in the Einstein-Straus spacetime is
governed mostly by partial transmissions and reflections. We recall
that these are produced by the most important features of the
potential $V_l$: a fairly well localized barrier near $r^* = 0$ and a
jump discontinuity at the boundary $\Sigma$. To explore the physics of
those transmissions and reflections, we consider a toy model involving
a simpler potential that still captures the essential features of the
original potential. The new wave equation is sufficiently simple that
we will be able to find an analytical solution. 

\subsection{Formulation of the toy model}   

The toy model is based on the wave equation 
\begin{equation}
\biggl( -\frac{\partial^2}{\partial \eta^2} + 
\frac{\partial^2}{\partial \chi^2} - V \biggr) \psi = 0,  
\label{5.1}
\end{equation}
in which the potential is given by 
\begin{equation}
V(\eta,\chi) = \left\{ 
\begin{array}{cl}
0 & \qquad 0 < \chi < \chi_0 \\
-2/\eta^2 & \qquad \chi > \chi_0 
\end{array}
\right. .
\label{5.2}
\end{equation}
The potential vanishes on the left-hand side of the boundary at $\chi
= \chi_0$, and the wave propagates freely in the interval $0 < \chi <
\chi_0$. This interval is our model for the region between the
potential barrier near $r^* = 0$ and the boundary $\Sigma$: we
effectively ``switch off'' the potential in the Schwarzschild region 
of the spacetime. The partial reflections off the potential barrier
are replaced by total reflections at $\chi = 0$, where we impose 
the boundary condition 
\begin{equation} 
\psi(\eta,\chi = 0) = 0. 
\label{5.3}
\end{equation}
On the right-hand side of the boundary, for $\chi > \chi_0$, the
potential is $\eta$-dependent, a behavior that is inherited by taking
the special case $l = 0$, $\xi = 0$, and $2a(\eta)/C = \frac{1}{2}
\eta^2$ of Eq.~(\ref{3.7}). For $\chi > \chi_0$, therefore,
Eq.~(\ref{5.1}) describes a $l = 0$, minimally-coupled wave
propagating in a spatially-flat cosmology. Notice that for $\chi <
\chi_0$, Eq.~(\ref{5.1}) describes a $l = 0$ wave propagating in flat
spacetime. 

The toy model incorporates an infinite potential barrier at $\chi =
0$ and a jump discontinuity at $\chi = \chi_0$. We believe that this
is sufficient to capture, in the simplest possible way, the essential
features of wave propagation in the Einstein-Straus spacetime. The toy
model, however, does not incorporate an event horizon, and it does not
take into account the effect of the $l(l+1)$ part of the potential. 

In the rest of the section we will integrate Eq.~(\ref{5.1})
analytically, subjected to the initial conditions 
\begin{equation}
\psi(u,v_0) = f(u), \qquad
\psi(u_0,v) = 0, 
\label{4.4}
\end{equation}
where $f(u)$ is a specified function with support in the
interval $v_0 < u < u_0$; this corresponds to an initially 
{\it outgoing} wave packet. We have re-introduced the null coordinates
$u$ and $v$, related to the original coordinates by $u = \eta - \chi$,
$v = \eta + \chi$. We will be interested mostly in the behavior of the
wave function at $\chi = \chi_0$. We sketch the evolution of the wave
packet in Fig.~13.  

\begin{figure}
\special{hscale=36 vscale=36 hoffset=-20.0 voffset=20.0
         angle=-90.0 psfile=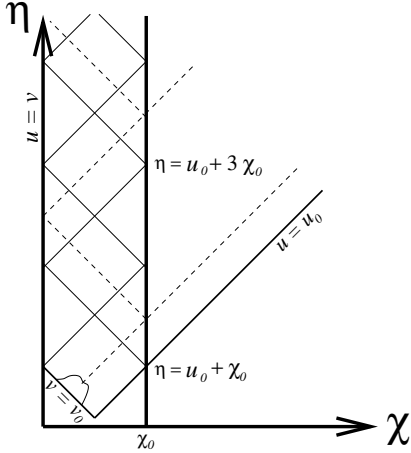}
\vspace*{2.5in}
\caption{Wave propagation in the toy model. The diagram sketches the 
         evolution of the initial wave packet, which is placed on the
         line $v = v_0$. When it first reaches the boundary at $\chi = 
         \chi_0$, the wave is partially transmitted and reflected. The
         reflected wave moves back toward $\chi = 0$, where it is
         totally reflected. The new outgoing wave encounters once more 
         the boundary, and it is again partially transmitted and
         reflected. The entire cycle repeats {\it ad infinitum}. The 
         first epoch of wave propagation refers to the wave's first
         encounter with the boundary; it begins at $\eta = u_0 +
         \chi_0$ and ends at $\eta = u_3 + 3\chi_0$. The second epoch
         refers to the wave's second encounter with the boundary; it
         begins at $\eta = u_0 + 3\chi_0$ and ends at $\eta = u_3 +
         5\chi_0$.}   
\end{figure} 

\subsection{Elementary solutions} 

We shall impose that $\psi$ and its derivative with respect to $\chi$
be continuous at $\chi = \chi_0$. This will allow us to describe the
partial transmissions and reflections that take place at the
boundary. The total reflections occurring at $\chi = 0$ are described
by Eq.~(\ref{5.3}). The  difficulty of the forthcoming calculations
resides entirely with the imposition of these matching
conditions. Solving the wave equation separately in the two domains
$\chi < \chi_0$ and $\chi > \chi_0$ is, by contrast, quite
trivial. For $V = 0$, the general solution to
Eq.~(\ref{5.1}) is 
\begin{equation}
\psi_{\rm left} = a(u) + b(v),
\label{5.5}
\end{equation}
in which $a$ and $b$ are arbitrary functions of their arguments. For
$V = -2/\eta^2$, the general solution to Eq.~(\ref{5.1}) is 
\begin{equation}
\psi_{\rm right} = \biggl( \frac{d}{du} - \frac{1}{\eta} \biggr) A(u) 
+ \biggl( \frac{d}{dv} - \frac{1}{\eta} \biggr) B(v), 
\label{5.6}
\end{equation} 
in which $A$ and $B$ are arbitrary functions of their arguments. 

We note in passing that Eqs.~(\ref{5.5}) and (\ref{5.6}) can easily 
be generalized to potentials that include a $l(l+1)/\chi^2$
term. Starting from a monopole solution $\psi_0$, a dipole ($l=1$)
solution is obtained by letting $\psi_1 = (-\partial/\partial \chi +
1/\chi) \psi_0$, and a quadrupole ($l=2$) solution is obtained by
letting $\psi_2 = (-\partial/\partial \chi + 2/\chi) \psi_1$. The
general rule is that a ($l+1$)-pole solution is obtained from a
$l$-pole solution by letting $\psi_{l+1} = [- \partial/\partial \chi 
+ (l+1)/\chi] \psi_l$.
 
\subsection{Matching: First epoch} 

We begin by matching the solutions (\ref{5.5}) and (\ref{5.6}) in the
interval $u_0 + \chi_0 < \eta < u_0 + 3\chi_0$. During this {\it first
epoch}, the following events take place: (i) the initially outgoing
wave packet represented by $f(u)$ arrives at $\chi = \chi_0$ and is
partially transmitted and reflected; (ii) the reflected wave moves
back toward $\chi = 0$ and is totally reflected there; (iii) the new
outgoing wave, which later will be represented by the function 
$f_{\rm new}(u)$, starts to move toward $\chi = \chi_0$. The first
epoch, therefore, includes only the first interaction between the
outgoing wave and the boundary.   

During the first epoch, 
\begin{equation}   
\psi_{\rm left} = f(u) + g(v),
\label{5.7}
\end{equation}
where $f(u)$ represents the initially outgoing wave packet, and $g(v)$
its partial reflection off the boundary; we have $g(v) = 0$ for $v <
u_0 + 2\chi_0$. On the other hand, 
\begin{equation}
\psi_{\rm right} = \dot{F}(u) - \frac{1}{\eta} F(u), 
\label{5.8}
\end{equation}
and this represents the wave that is partially transmitted across the
boundary; an overdot indicates differentiation with respect to
$u$. The functions $g(v)$ and $F(u)$ are determined by imposing
continuity of $\psi$ and $\psi_{,\chi}$ across $\chi = \chi_0$. The
equation that determines $F(u)$ is 
\begin{equation}
\ddot{F}(u) - \frac{\dot{F}(u)}{u+\chi_0} + \frac{F(u)}{2(u+\chi_0)^2} 
= \dot{f}(u). 
\label{5.9}
\end{equation}
The equation that determines $g(v)$ is 
\begin{equation}
g(v) = -f(v-2\chi_0) + \dot{F}(v-2\chi_0) 
- \frac{F(v-2\chi_0)}{v-\chi_0}.   
\label{5.10}
\end{equation} 

We must first integrate Eq.~(\ref{5.9}) subjected to the initial
conditions $F(u_0) = \dot{F}(u_0) = 0$. Standard integration
techniques \cite{24} reveal that the required solution is    
\begin{eqnarray}
F(u) &=& \frac{1}{2} (u+\chi_0)^{1+\gamma} \int_{u_0}^u
\frac{f(u')}{(u'+\chi_0)^{1+\gamma}}\, du' 
\nonumber \\ & & \mbox{}
+ \frac{1}{2} (u+\chi_0)^{1-\gamma} \int_{u_0}^u
\frac{f(u')}{(u'+\chi_0)^{1-\gamma}}\, du', 
\label{5.11}
\end{eqnarray}
where $\gamma \equiv 1/\sqrt{2}$. Substituting this into
Eq.~(\ref{5.10}) gives
\begin{eqnarray}
g(v) &=& \frac{\gamma}{2} (v-\chi_0)^\gamma \int_{u_0 + 2\chi_0}^v
\frac{f(v'-2\chi_0)}{(v'-\chi_0)^{1+\gamma}}\, dv' 
\nonumber \\ & & \mbox{}
- \frac{\gamma}{2} (v-\chi_0)^{-\gamma} \int_{u_0 + 2\chi_0}^v 
\frac{f(v'-2\chi_0)}{(v'-\chi_0)^{1-\gamma}}\, dv'. 
\label{5.12}
\end{eqnarray} 
We recall that this ingoing wave is totally reflected at $\chi = 0$,
and that it becomes the new outgoing wave $f_{\rm new}(u) = -g(v)$.  

By substituting Eq.~(\ref{5.11}) into Eq.~(\ref{5.8}) or
Eq.~(\ref{5.12}) into Eq.~(\ref{5.7}), we finally obtain the wave
function $\psi(u,v)$. Its expression on the boundary is   
\begin{eqnarray} 
\psi(\eta,\chi_0) &=& f(\eta-\chi_0) + \frac{\gamma}{2} \eta^\gamma  
\int_{u_0 + \chi_0}^\eta 
\frac{f(\eta'-\chi_0)}{\eta^{\prime 1+\gamma}}\, d\eta' 
\nonumber \\ & & \mbox{}
- \frac{\gamma}{2} \eta^{-\gamma}  
\int_{u_0 + \chi_0}^\eta 
\frac{f(\eta'-\chi_0)}{\eta^{\prime 1-\gamma}}\, d\eta'. 
\label{5.13}
\end{eqnarray}
This expression is valid in the interval $u_0 + \chi_0 < \eta <u_0 +
3\chi_0$ only.  

\subsection{Matching: Second epoch} 

We continue our integration of Eq.~(\ref{5.1}) and carry out the
matching of Eqs.~(\ref{5.5}) and (\ref{5.6}) in the interval $u_0 +
3\chi_0 < \eta < u_0 + 5\chi_0$. During this {\it second epoch}, the
following events take place: (i) the new outgoing wave 
$f_{\rm  new}(u)$ arrives at $\chi = \chi_0$ and is partially
transmitted and reflected; (ii) the reflected wave moves back toward
$\chi = 0$ and is totally reflected there; (iii) the new outgoing wave
starts to move toward $\chi = \chi_0$. 

During the second epoch, 
\begin{equation}   
\psi_{\rm left} = f_{\rm new}(u) + g_{\rm new}(v), 
\label{5.14}
\end{equation}
where  
\begin{eqnarray}
f_{\rm new}(u) &=& -\frac{\gamma}{2} (u-\chi_0)^\gamma \int_{u_0 +
  2\chi_0}^u \frac{f(u'-2\chi_0)}{(u'-\chi_0)^{1+\gamma}}\, du'  
\nonumber \\ & & \mbox{}
+ \frac{\gamma}{2} (u-\chi_0)^{-\gamma} \int_{u_0 + 2\chi_0}^u 
\frac{f(u'-2\chi_0)}{(u'-\chi_0)^{1-\gamma}}\, du' 
\nonumber \\ & &
\label{5.15}
\end{eqnarray} 
represents the new outgoing wave, and $g_{\rm new}(v)$ its partial
reflection off the boundary. On the other hand,  
\begin{equation}
\psi_{\rm right} = \dot{F}(u) - \frac{1}{\eta} F(u) 
\label{5.16}
\end{equation}
represents the partially transmitted wave; because $F(u)$ is 
continuous at $u = u_0 + 2\chi_0$, there is no need to distinguish the
second-epoch $F(u)$ with a label ``new''. 

The functions $F(u)$ and $g_{\rm new}(v)$ are determined by
Eqs.~(\ref{5.9}) and (\ref{5.10}), in which we replace $f(u)$ by
$f_{\rm new}(u)$ and $g(v)$ by $g_{\rm new}(v)$. Equation (\ref{5.9})
is solved under the conditions that $F(u_0 + 2\chi_0)$ and
$\dot{F}(u_0 + 2\chi_0)$ match the values obtained from
Eq.~(\ref{5.11}). This gives 
\begin{eqnarray}
F(u) &=& \frac{1}{2} (u+\chi_0)^{1+\gamma} \Biggl[ 
\int_{u_0}^{u_0+2\chi_0} \frac{f(u')}{(u'+\chi_0)^{1+\gamma}}\, du'  
\nonumber \\ & & \mbox{}
+ \int_{u_0+2\chi_0}^u 
  \frac{f_{\rm new}(u')}{(u'+\chi_0)^{1+\gamma}}\, du' \Biggr]  
\nonumber \\ & & \mbox{}
+ \frac{1}{2} (u+\chi_0)^{1-\gamma} \Biggl[ 
  \int_{u_0}^{u_0+2\chi_0} \frac{f(u')}{(u'+\chi_0)^{1-\gamma}}\, du'   
\nonumber \\ & & \mbox{}
+ \int_{u_0+2\chi_0}^u 
  \frac{f_{\rm new}(u')}{(u'+\chi_0)^{1-\gamma}}\, du' \Biggr], 
\label{5.17}
\end{eqnarray}
and $g_{\rm new}(v)$ can be obtained immediately from
Eq.~(\ref{5.10}); because this result is not needed for further
calculations, we shall not display it here. Substituting
Eq.~(\ref{5.17}) into Eq.~(\ref{5.16}) and evaluating on the boundary,
we obtain 
\begin{eqnarray} 
\psi(\eta,\chi_0) &=& f_{\rm new}(\eta-\chi_0) 
+ \frac{\gamma}{2} \eta^\gamma \Biggl[ 
  \int_{u_0 + \chi_0}^{u_0 + 3\chi_0} 
  \frac{f(\eta'-\chi_0)}{\eta^{\prime 1+\gamma}}\, d\eta' 
\nonumber \\ & & \mbox{}
+ \int_{u_0 + 3\chi_0}^{\eta} 
  \frac{f_{\rm new}(\eta'-\chi_0)}{\eta^{\prime 1+\gamma}}\, d\eta' 
\Biggr] 
\nonumber \\ & & \mbox{}
- \frac{\gamma}{2} \eta^{-\gamma} \Biggl[ 
  \int_{u_0 + \chi_0}^{u_0 + 3\chi_0} 
  \frac{f(\eta'-\chi_0)}{\eta^{\prime 1-\gamma}}\, d\eta' 
\nonumber \\ & & \mbox{}
+ \int_{u_0 + 3\chi_0}^{\eta} 
  \frac{f_{\rm new}(\eta'-\chi_0)}{\eta^{\prime 1-\gamma}}\, d\eta' 
\Biggr]. 
\label{5.18}
\end{eqnarray}
This expression is valid in the interval $u_0 + 3\chi_0 < \eta <u_0 +
5\chi_0$ only. 

\begin{figure}
\special{hscale=35 vscale=35 hoffset=-20.0 voffset=10.0
         angle=-90.0 psfile=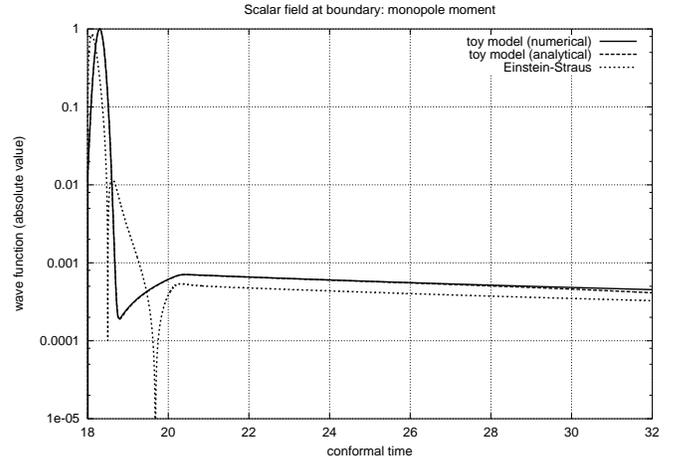}
\vspace*{2.6in}
\caption{A plot of the absolute value of the wave function on the
         boundary, as a function of conformal time $\eta$. The solid
         curve is obtained from a numerical integration of
         Eq.~(\ref{5.1}), starting with the gaussian wave packet
         described by Eq.~(\ref{4.2}). For this integration we set
         $u_0 = 17$, $v_0 = 17$, $v_c = 17.3$, $\sigma = 0.1$, and
         $\chi_0 = 1$. The dashed curve is obtained from the
         analytical results of Eqs.~(\ref{5.13}) and (\ref{5.18}), in
         which we substitute the totally reflected gaussian wave
         packet for $f(u)$. The two curves labelled ``toy model''
         start to diverge at the end of the second epoch, when $\eta =
         u_0 + 5\chi_0 = 22$. The dotted curve represents a monopole
         wave in the Einstein-Strauss spacetime, which displays a
         similar late-time behavior.}      
\end{figure} 

\subsection{Wave propagation in the toy model} 

Equation (\ref{5.13}) gives $\psi(\eta,\chi_0)$ in the interval $u_0 +
\chi_0 < \eta < u_0 + 3\chi_0$, and Eq.~(\ref{5.18}) does the same in
the interval $u_0 + 3\chi_0 < \eta < u_0 + 5\chi_0$. It can be checked
that these values are continuous at $\eta = u_0 + 3\chi_0$. While
integration of Eq.~(\ref{5.1}) could easily be carried out beyond the
second epoch by repeated application of the general method, we shall
stop here: to continue beyond this point would give rise to messy
expressions that would not teach us anything new. The message of
Eqs.~(\ref{5.13}) and (\ref{5.18}) is already clear: The multiple 
transmissions and reflections that take place at $\chi = \chi_0$
produce a wave function with a fairly complicated mathematical
description, and its late-time behavior is not well described by a
simple inverse power-law.  

In Fig.~14 we compare the wave function obtained by integrating
Eq.~(\ref{5.1}) to a $l=0$ wave in the Einstein-Straus
spacetime. The figure reveals that both waves have a similar late-time 
behavior, providing support for our claim that wave propagation in
the Einstein-Straus spacetime is governed mostly by partial
transmissions and reflections.     

\section*{Acknowledgments}

This work was supported by the Natural Sciences and Engineering
Research Council of Canada.

\end{document}